\documentclass[manuscript]{aastex}

\slugcomment{Submitted to ApJ}

\shorttitle{Global High-resolution $N$-body Simulation of Planet Formation}
\shortauthors{Kominami et al.}

\begin{document}


\title{Global High-resolution $N$-body Simulation of Planet Formation I. \\
    Planetesimal Driven Migration}


\author{J. D. Kominami\altaffilmark{1}}
\affil{Earth-Life Science Institute, Tokyo Institute of Technology}
\email{kominami@mail.jmlab.jp}

\author{H. Daisaka\altaffilmark{2}}
\affil{Hitotsubashi University}
\email{daisaka@phys.science.hit-u.ac.jp}

\author{J. Makino\altaffilmark{3,1}}
\affil{RIKEN Advanced Institute for Computational Science}
\email{makino@mail.jmlab.jp}

\and

\author{M. Fujimoto\altaffilmark{4}}
\affil{Japan Aerospace Exploration Agency}
\email{fujimoto.masaki@jaxa.jp}


\altaffiltext{1}{Earth-Life Science Institute, Tokyo Institute of Technology}
\altaffiltext{2}{Hitotsubashi University}
\altaffiltext{3}{RIKEN Advanced Institute for Computational Science}
\altaffiltext{4}{Japan Aerospace Exploration Agency}


\begin{abstract}
We investigated whether outward Planetesimal Driven Migration (PDM)
takes place or not in simulations 
when the self gravity of planetesimals is included.
We performed $N$-body simulations of planetesimal disks with large width
(0.7 - 4AU) which ranges over the ice line.
The simulations consisted of two stages.
The first stage simulations were carried out to see
the runaway growth phase using the planetesimals of initially the
same mass.
The runaway growth took place both at the inner edge of the disk and at the
region just outside the ice line.
This result was utilized for
the initial setup of the second stage simulations
in which the runaway bodies just outside the ice line were replaced by
the protoplanets with about the isolation mass.
In the second stage simulations, the outward migration
of the protoplanet was followed by the stopping of the migration due to
the increase of the random velocity of the planetesimals.
Due to this increase of random velocities, one of the PDM criteria derived
in Minton and Levison (2014) was broken.
In the current simulations, the effect of the gas disk is not considered.
It is likely that the gas disk plays an important role in
planetesimal driven migration, and we plan to study its effect
in future papers.

\end{abstract}





\section{Introduction}

In the current planet formation theory, the formation of the
terrestrial planets and gas giant cores have been explained in the following 
way (e.g. Morbidelli et al. 2012).
First, the runaway accretion of the planetesimals takes place and the
runaway bodies, sometimes called embryos or protoplanets, are formed
(Wetherill \& Stewart 1989, Kokubo \& Ida 1996).
They heat up the random velocities of surrounding planetesimals.
This increase of random velocities causes the slowing down of the growth
rate of both protoplanets themselves and the surrounding planetesimals.
Thus, multiple protoplanets
are formed and they grow keeping their distance to $\sim$ 10 Hill's radius.
This phase is called
the oligarchic phase (Kokubo \& Ida 1998) and it continues until
their mass reaches the isolation
mass.  The isolation mass is about the Mars mass in the terrestrial
planet region.  After most of the disk gas has been dissipated,
the orbits of protoplanets cross with each other.
As a result, protoplanets collide with each other to form terrestrial
planets.

This scenario has been investigated using $N$-body 
simulations which included the self-gravity of the planetesimals:
Runaway growth (e.g. Kokubo \& Ida 1996), oligarchic growth
(Kokubo \& Ida 1998),
and the formation of planets form protoplanets through giant impact phase
(e.g. Kominami \& Ida 2002).

In these simulations, the radial width of the planetesimal disk ranged from
$\sim 0.02$ to $\sim 1.0$AU which is narrow compared with
the size of solar system ($\sim 40$ AU).
One of the reasons why the planetesimal disks of previous simulations
were narrow was the limitation of a number of particles could be used.
For example, the number of particles used in Kokubo \& Ida (2002) was
about 10000.
If larger number of particles were used,
the simulating time would not be practical.

Although such simulations of narrow disks had been useful to study
individual process,
it is important to study wider planetesimal disks, at least
ranging from terrestrial planet region to the Jovian planet region.
For instance, there is the ice line at which the solid surface density 
increases
by a factor of a few.  It would affect the growth of planetesimals.
Water delivery from outside
the ice line to the inner region of the ice line has been considered
(Walsh et al. 2011, Raymond et al. 2012).
Migration mechanisms of planets such as type-I migration 
(Tanaka et al. 2002)
and planetesimal driven migration (Ida et al 2000, 
Minton \& Levison 2014)
have crucial effect on the formation of the planetary systems.
In order to understand planetary formation processes which include
such mechanisms, one needs to consider the disks extending
$\sim 0.5$AU to $\sim 10$AU.
Such simulations of disks with wide radial range
requires the number of particles
such as several times $10^5$.
We have developed a parallel $N$-body code which enables us to simulate
several times $10^5$ particles for $10^5$ years in a reasonable 
time (Kominami et al. 2016).
Our ultimate goal is to follow the planetary formation process from
planetesimals to final planets directly with a single $N$-body
simulation which includes the effects of gas drag, type-I migration
and collisional fragmentation, to understand how terrestrial and
outer planets are formed.
In the forthcoming papers, we will study the effects of gas
drag, type-I migration and collisional fragmentation, step by step.

In this paper, we focus on the mechanism called planetesimal driven migration
(PDM, Kirsch et al. 2009, Ormel et al. 2012, Minton \& Levison 2014)
which can move the planets outward in a planetesimal disk.
PDM is triggered by the formation of asymmetry in the density structure
in the planetesimal disk
around the protoplanets.  The asymmetric density structure results in
the asymmetric net torque on the protoplanets.
The torque makes the protoplanet migrate to one direction.
On the plane of semimajor axis and the random velocity,
individual planetesimals
within the horseshoe region in the radially outside are scattered to the
region radially inside when a protoplanet moves outward, transporting
angular momentum to the protoplanet.
PDM has been studied semi-analytically
(Ida et al. 2000, Kirsch et al 2009, Minton \& Levison 2014).
Minton and Levison (2014, hereafter ML14) also studied PDM using
test particle simulations, in which the interaction between the large
protoplanets and surrounding planetesimals are included but interactions
between planetesimals are neglected.
They derived criteria for PDM to take place.
The criteria are (1) the mass ratio criterion, (2) the mass resolution
criterion, (3) the disk eccentricity criterion,
(4) the crowded criterion and (5) the growth timescale criterion.
In order for PDM to start, the criterion (2) indicates that
the mass of surrounding planetesimals has to be
small ($\sim 1/100$) compared to the mass of the protoplanet.
If we consider a Mars-sized protoplanet, individual planetesimals has
to be $\sim 10^{24}$g or less, and the number of planetesimals is
several times $10^5$ if we consider a region of several AU.

Although the effect of the self gravity is not considered in previous studies,
it might play an important role in the PDM
process because of the following reasons.  As shown in criterion (3)
of ML14, the velocity dispersion of the
planetesimals should be small enough for PDM to take place and to
continue.
It is expected that this decrease of angular momentum might 
eventually halt PDM.
The self gravity might change the way PDM works.
Thus, we believe it is important to study the PDM process using
the simulations with the self gravity of planetesimals.

In this paper, we performed simulations in
two-stages.
In the first stage simulations, we started with equal mass planetesimals
distributed in a disk extending from 0.7 AU to 4.0 AU with $82362$
particles.
Runaway growth took place at the inner edge and at right outside the
ice line.
Since we did not include the effect of the gas drag,
the velocity dispersion of planetesimals would increase
to unrealistic values if we continued the simulation.
In order to save the simulation time and to avoid the unrealistic increase
of the velocity dispersion,
we introduced the assumption that
the two largest protoplanets grow to the size of
about the isolation mass.
After we increased the mass of these two protoplanets,
we restarted the simulation as the second stage simulation.

In the second stage simulations, the outer protoplanet,
which was initially at $\sim 2.3$AU,
moved outward and reached $\sim 3$AU.
The viscous stirring among the planetesimals increased the
random velocity of the planetesimals and the outward migration was eventually
halted.
To our knowledge, this is the first $N$-body simulation with
self-gravity of the planetesimals in which PDM took place.

One of the critical issues of the planetary accretion theory is the
formation timescale of the cores of ice giants in our Solar system.
The in-situ formation timescale of the core of Neptune is estimated
to be longer than the age of the Solar system (Kokubo \& Ida 2002).
We can expect that
PDM would solve this issue, since it enables the
core formation of gas giants and icy planets in the inner
region of the disk.

In section 2, we describe the method of the simulation.
The result of the first stage of the simulation is presented in section 3,
and that of the second stage is in section 4.
Section 5 is devoted to discussion and summary.


\section{Numerical Method}
For our $N$-body simulations, we used
Kninja (Kominami et al. 2016), 
which is a parallelized $N$-body simulation code developed for 
K computer.
It uses
the 4th order Hermite scheme (Makino \& Aarseth 1992) with 
block timestep and the two-dimensional Ninja algorithm 
(Nitadori et al. 2006)  
which provides good load balance between cpus
and excellent scalability.
Perfect accretion is assumed for 
collisions between the particles. 
Typically, in our simulations, 
the relative energy error is $\sim 10^{-6}$ after integration of $10^5$ years.
Since we include the planetesimal-planetesimal gravitational interaction,
there is no approximation for 
gravitational interaction between the particles in our simulations.
In addition, since the scheme we use is the 4th order Hermite scheme,
we can achieve sufficient accuracy for long term orbital evolution.
Second order integrators are not sufficient for such
simulations (Carter et al. 2015).

Our simulations consist of two stages.
First stage simulations start with equal mass planetesimals.
The physical density of the planetesimals and the protoplanets 
is $3{\rm g}{\hspace{0.5em}}{\rm cm}^{-3}$.
Using the results of the first stage simulations, we constructed
the initial conditions of the second stage simulations.
The two of the largest runaway bodies outside the ice line 
which was formed in the first stage simulations 
were replaced by the protoplanets.
The summary of simulations is listed in table 1.
Since the time to simulate $82362$ planetesimals for $5\times 10^4$ years
using 1024 nodes was about 10 days including the waiting period, 
the computational (wall clock) time for the runaway bodies to grow to
the size of $0.1M_\oplus$ without gas drag would be about several months. 

\section{First Stage Simulations : until the Runaway Growth}
\subsection{Setup for the First Stage Simulations}

Our setup of the first stage simulations is almost the same 
as those used in 
Kokubo \& Ida (2002) except that we extended the planetesimal disk
so that it includes the region outside the ice line.  
We distributed planetesimals with equal mass of $m=10^{24}$g from
0.7 AU to 4.0 AU.
The distribution of planetesimals in radial direction obeys
the solid surface density of Minimum Mass Solar Nebula (MMSN) 
which is given by (Hayashi et al. 1985)
\begin{eqnarray}
\Sigma_{\rm d} &=& 7.1 \times \left( \frac{r}{{\rm 1AU}} \right)^{-3/2}
[{\rm g}{\hspace{3pt}}{\rm cm}^{-2}] {\hspace{2em}} (r<{\rm ice}{\hspace{5pt}}{\rm line}) , \\
\Sigma_{\rm d} &=& 30 \times \left( \frac{r}{{\rm 1AU}} \right)^{-3/2}
[{\rm g}{\hspace{3pt}}{\rm cm}^{-2}] {\hspace{2em}} (r>{\rm ice}{\hspace{5pt}}{\rm line}).
\end{eqnarray}
Thus, the number density of the planetesimals
is larger outside the ice line.
Initial distribution of the eccentricity($e$) and the inclination($i$) of 
the planetesimals are given by the Rayleigh distribution with 
\begin{equation}
\langle e^2 \rangle^{1/2} = 2\langle i^2 \rangle^{1/2} = 2.0 h,
\end{equation}
where $h$ is the reduced Hill radius $h=(m/3M_\odot)^{1/3}$ 
(Kokubo \& Ida 1998).
The initial number of planetesimals in the disk is 82362.
We carried out two simulations with 
different initial condition of particles (position and velocity), 
which are generated with different seeds of random numbers (S1a and S1b).
We used the radius enhancement factor, $f$, 
which is used to accelerate the growth of planetesimals
(Kokubo \& Ida 1996) in this first stage simulations.  
We used $f=3$ in the first stage simulations to save the computational time.
Note that we did not use the enhancement factor in second stage 
simulations (i.e. $f=1$).

The ice line is at 2.7 AU in our present Solar system.
However, it was pointed out that the ice line might 
have been closer to the Sun 
due to the viscous accretion of the gas  and due to the stellar 
evolution of the central star (Oka et al. 2011).
Therefore, we placed the ice line at 2.0 AU in our simulations.

\subsection{Emergence of Runaway Bodies}
Figure \ref{fig.sim_1_e_a} shows the snapshots of run S1a 
on the plane of the 
semimajor axis (denoted by $a$) and the random velocity 
($\sqrt{e^2 + i^2 }$). 
The size of circles indicates the mass of a planetesimal.
When the mass of a planetesimal grows to more than ten times its
initial mass, we call it a runaway body. It is shown in a black
circle. As seen in figure \ref{fig.sim_1_e_a},
the runaway bodies are formed not only at the inner edge of the disk
but also at the region just outside the ice line ($\sim$ 2AU, described 
as the dashed line).
This is due to the fact that 
the increase of the solid surface density
at the ice line enhances the collision probability, 
which accelerates the growth of the planetesimals.
Meanwhile, such large bodies outside the ice line did not emerge in 
previous simulations without ice line (Kokubo \& Ida 2002).

Figure \ref{fig.RUNAWAY.INNER.OUTER} shows the mass evolution of the 
largest runaway bodies from 2.0 to 2.5 AU (solid curve)
and in the region from 0.7 to 1.0 AU (dotted curve).
The average mass excluding the largest one in each region
is shown as the dashed curve and in the dot-dashed
curve, respectively.
The mass ratio of the largest protoplanet 
to the average mass excluding the largest one becomes more that 100 at the end
of the simulation.

Here we estimate the growth timescale of the runaway protoplanets.
The growth timescale ($T_{\rm grow}$) for the planetesimal to reach
the mass of $M$ is empirically derived in Kokubo \& Ida (2002)
and written as
\begin{eqnarray}
T_{\rm grow} &\simeq& 4.8 \times 10^3 f^{-1} \langle \tilde{e}^2
\rangle \left(\frac{C}{2}\right)^{-1}
\left( \frac{M}{10^{26}{\rm g}} \right)^{1/3}
\left( \frac{\rho_{\rm p}}{2{\rm g}{\hspace{1em}}{\rm cm}^{-3}} \right)^{1/3} \nonumber \\
& &
\left( \frac{\Sigma_{\rm d}}{10{\rm g}{\hspace{1em}}{\rm cm}^{-2}} \right)^{-1}
\left( \frac{a}{1{\rm AU}} \right)^{1/2} 
\left( \frac{M_\ast}{M_\odot} \right)^{-1/6} {\rm years}
\label{eq1}
\end{eqnarray}
where $C$ is the accretion acceleration factor of a few that 
reflects the effects of the eccentricity and inclination distributions 
of planetesimals and the Solar gravity, $f$ is the radius enhancement
factor, $\rho_{\rm p}$ is the physical
density and $M_\ast$ is the mass of the central star.
The largest protoplanet outside the ice line of the last panel
of figure \ref{fig.sim_1_e_a} 
has the mass of $2.8 \times 10^{25}$g. 
In order to estimate the growth timescale of the runaway bodies
using eq.(\ref{eq1}),
we plot the root-mean-square (RMS) eccentricity evolution 
of the planetesimals (figure \ref{fig.ECC_INNER_OUTER}).
The RMS eccentricity of the planetesimals in 0.7 - 1.0 AU is plotted
in the solid curve, and in 2.0 -2.5 AU in the dotted curve.
Using the values in the solid curve, 
the growth time scale $T_{\rm grow}$ can be estimated as
$\sim 7\times 10^4$ years for 
$M=2.8 \times 10^{25}$g at $\sim 2$AU
if the radius enhancement factor is 3 and $C=2$.
In S1a, the runaway bodies outside the ice line 
grow to mass of $2.8 \times 10^{25}$g at $5.6\times 10^4$ years,
which is consistent with the result of previous works 
(Kokubo \& Ida 1996, 1998, 2002).

\section{Second Stage Simulations : Migration of the Protoplanets}
\subsection{Criteria for PDM}
ML14 listed five criteria that
have to be satisfied simultaneously in order for PDM to take place.
We briefly go over those criteria and discuss the importance of 
the self gravity of the planetesimals and how it would affect the
PDM.  
The criteria are the following. (1) The mass ratio criterion; 
the amount of planetesimals 
within $5h a_{\rm p}$ of the protoplanet at $a_{\rm p}$
should be larger than three times the protoplanet's mass $m_{\rm p}$.
(2) The mass resolution criterion; 
the mass ratio between the protoplanet and the planetesimal ($m_{\rm p}/m$) 
has to be larger than $\sim 100$ for the migration to be 
monotonic in one direction.
(3) The disk eccentricity criterion;
the eccentricity of the planetsimals surrounding the protoplanet
has to be less than $3h$ for the migration to take place.
(4) The crowded criterion; 
other growing bodies near the protoplanet can distract the migration.
(5) The growth timescale criterion; 
if a migrating body encounters with a protoplanet of similar size, 
the migration halts.
The migration rate depends strongly on the first three 
criteria.  By carrying out self-consistent $N$-body simulations,
the evolution of the random velocity and the mass of the planetesimals
can be followed.
Viscous stirring of the planetesimals can increase
the random velocity of the planetesimals. This increase might affect
the criterion (3).  
Criteria (1) and (2) may be altered if the growth of the planetesimals are
considered as well.  
Here we would like to see how first three criteria change 
when $N$-body simulations which include the self-gravity of the 
planetesimals are performed.

\subsection{Setup for the Second Stage Simulations}
Here we briefly describe how we set up the initial condition for the 
second stage simulations.
We replaced the largest two runaway bodies outside 2AU that
emerged in the first stage simulation with the mass of the protoplanet.
These bodies emerged at $5.6 \times 10^4$ years,
and reached a mass of $0.0047M_\oplus$.
We parametarized $m_{\rm proto}$ and investigated three different 
cases; $m_{\rm proto}=0.1M_\oplus$ (S2aM0.1), $0.3M\oplus$ (S2aM0.3) 
and $0.03M_\oplus$ (S2aM0.03).
As noted in section 3, we did not use the radius enhancement factor
in second stage simulations.

In run S2aM0.1,
the planetesimals within the distance of $5 r_{\rm hill}$ from
$m_{\rm proto}(=0.1M_\oplus)$ were removed from the system as shown in
figure \ref{fig.sim_2_initial_e_a}.  The Hill radius
$r_{\rm Hill}$ is written as
\begin{equation}
r_{\rm hill}=h_{\rm M}a_{\rm p}=(m_{\rm proto}/3M_\odot)^{1/3}a_{\rm p},
\end{equation} 
where $h_{\rm M}$ is the reduced Hill radius of the protoplanet
and $a_{\rm p}$ is the semimajor axis of the protoplanet 
(Ida 1990, Kokubo \& Ida 1998).
In order to see only the effect of the different $m_{\rm proto}$,
we uses the same distribution of planetesimals for runs S2aM0.1, 
S2aM0.3 and S2aM0.03.

Following is the justification for replacing the runaway bodies outside 2AU
with $m_{\rm proto}$.
In the gas-free case, if we wait for the runaway bodies to grow to 
the mass of $0.1 M_\oplus$, it would take $\sim 7 \times 10^5$ years
(Kokubo \& Ida 2002).  The eccentricity of the planetesimals would 
increase to $\sim 10h$
(Ida 1990, Kokubo \& Ida 1998).
However, such increase of the random velocity 
is not realistic if the gas drag is taken into account.
In order to avoid the unreasonable increase of the random 
velocity of the planetesimals,
we replaced the runaway bodies with $m_{\rm proto}$ in early phase
when the random velocity of the planetesimals were not significantly 
increased.
If the gas drag is included, the eccentricity of the planetesimals 
would be about $6h$
(Kokubo \& Ida 2002) which is about the same value as that at the 
moment we increased the mass of runaway bodies.

The total mass of the planetesimals removed is 0.18 $M_\oplus$
and the total mass of the two new protoplanets is 0.2 $M_\oplus$.
Thus, the total mass is approximately conserved in the case of 
S2aM0.1 and S2bM0.1.

\subsection{Radial Migration of the protoplanets}
Figure \ref{fig.sim_2_e_a}
shows the snapshots of run 
S2aM0.1 on the plane of $a$ and $\sqrt{e^2+i^2}$.
The protoplanets with initial mass of $m_{\rm proto}$
(in this case $m_{\rm proto}=0.1M_\oplus$) are 
shown in black circles.
The outer protoplanet 
moves outward to 2.7AU and the inner protoplanet
moves inward to 2AU.  The displacement of the outer protoplanet is 
$0.4$AU.  After the 
protoplanet reaches $2.7$AU, it moves back a little.

The random velocity of the planetesimals surrounding 
the protoplanet increases as shown in figure \ref{fig.sim_2_e_a}.
The planetesimals with Jacobi energy close to 0 are scattered in
V-like direction around the protoplanet (Nakazawa \& Ida 1988).
This increase shows that 
the protoplanet stirs the surrounding planetesimals as in 
Ida \& Makino (1993).
The stirring of the planetesimals into the gap that is formed in
the planetesimal disk serves as a ``kick'' for the PDM to start.

Figure \ref{fig.sim_4_e_a} shows the case of run S2aM0.3. 
Outward migration is apparent.
In this case, the mass ratio between the protoplanet
and a planetesimal is $\sim 300$.  
The migration of the outer protoplanet was from 2.45AU
to 2.65AU. The migration speed
is slow compared to that in the case of run S2aM0.1.
Figure \ref{fig.sim_3_e_a} shows the case of $m_{\rm proto}=0.03M_\oplus$
(S2aM0.03). The displacement of the protoplanet is small.
The outer protoplanet stays at 2.45 AU.
The inner protoplanet moves inward a little bit from 2.2AU to 2.1AU.

The evolution of the semimajor axis of the outer protoplanets
in runs S2aM0.03, S2aM0.1 and S2aM0.3 is 
shown in figure \ref{fig.sim_2_sa_out}.
Constant migration is observed in runs S2aM0.1 and S2aM0.3,
although the migration rate is lower in run S2aM0.3.
In the case of S2aM0.03, the radial migration is very small,
except for the first 
$\sim 8000$ years.

The migration rate from Kirsh et al. (2009) is about $10^{-6}$AU/year
if our parameters are used.  Our simulations resulted in somewhat faster 
migration in some cases.  The reason may be the size of the kick and the 
stochastic nature of PDM.  This is one of the reasons we need to carry
out more simulations to obtain statistically reliable results.

In addition to run S2aM0.1, we performed run S2bM0.1
in order to investigate how the results depends on the initial 
random seed.
We utilized the outcome of run S1b to generate the initial condition of 
S2bM0.1.
Figure \ref{fig.sim2_evo_2sample} shows the evolution of the 
semimajor axis of the 
outer protoplanets for runs S2aM0.1 and
S2bM0.1.  
The migration continued up to $3.1$AU in run S2bM0.1.
The amount of migration can change by about $0.5$AU just 
because of random seeds.  
We need to carry out more simulations
to obtain statistically reliable results.

\subsection{Comparison with Criteria of ML14}
In this section, in order to investigate the criteria for PDM
to start/halt, we checked 
whether the criteria of ML14 are satisfied or not in our simulations.
We actually looked at the first three criteria only, since the last two are 
related to the cases where there are multiple outgoing protoplanets.

Criterion (1) of ML14 tells us that the amount of planetesimals 
within $5r_{\rm hill}$ outside the protoplanet at semimajor
axis $a_{\rm p}$ 
should be larger than three times the mass of the protoplanet 
$m_{\rm p}$ in order for the monotonic migration to take place.
To see whether criterion (1) is satisfied or not, we plot  
the total mass of planetesimals in front of the outgoing protoplanet
in figure \ref{fig.sim2_proto_mass.new}.  
Here we define the total mass of planetesimals ($m_{\rm A}$) with
the semimajor axis of $a_{\rm p}<a<a_{\rm p}+5r_{\rm hill}$ 
outside the outgoing protoplanet.
The reason we incorporated $5r_{\rm hill}$ is because 
the region in which the planetesimals are scattered in the horseshoe 
region is around $5r_{\rm hill}$ of the protoplanet (Nakazawa \& Ida 1988).
The vertical axis 
in figure \ref{fig.sim2_proto_mass.new} is $m_{\rm A}$ 
normalized by $m_{\rm proto}$.  
The thin line of $m_{\rm A}/m_{\rm proto}=1/3$ shows the 
criterion (1) of ML14.
All of our simulations resulted in the value above the line.
Hence, the criteria (1) is always satisfied in our simulations.

Criterion (2) of ML14 states that the mass ratio between the 
protoplanet and the planetesimal ($m_{\rm p}/m$) 
has to be larger than $\sim 100$ for the migration to be 
monotonic in one direction.
In order to compare our results with ML14, we plot the time evolution
of the averaged planetesimal mass 
normalized by $m_{\rm proto}$ 
within $5 r_{\rm hill}$ outside of the outer protoplanet in 
figure \ref{fig.sim2_proto_avg.new}.
Criterion (2) of ML14 is satisfied in all three runs, and the mass ratio becomes
large in all runs due to the growth of the protoplanet. 
Thus, criterion (2) is always satisfied in our
simulations.

Criterion (3) of ML14 states that the random velocity of
the planetesimals should be less than $\sim 3 h\simeq0.014$.
In order to see the random velocity evolution of the planetesimals 
in our simulations, we plot the eccentricity evolution (RMS of the 
eccentricity) of the planetesimals within $5 r_{\rm hill}$ outside 
the outward migrating protoplanet in figure \ref{fig.eccevo_all.new}. 
The migration stops when the eccentricity of the planetesimals
exceeds the value of $5 - 6h$ in all four cases
(figure \ref{fig.sim_2_sa_out}).
Thus, as far as our runs are concerned, criterion (3) of ML14,
namely the disk eccentricity criterion determines if the PDM continues
or not.

As seen in figure \ref{fig.sim_2_e_a}, the inner protoplanet 
loses the angular momentum throwing the
planetesimals from the inner region to the outer region
in the same manner as the outer protoplanet moves outward throwing 
the planetesimals from the outer region to the inner region.
The inward migration stops at the ice line probably due to the decrease
of the number of planetesimals in the disk.
We will investigate the inward movement further in the next paper.

\section{Summary and Discussion}
\subsection{Summary of $N$-body Simulations}
{In this paper, we carried out high resolution $N$-body simulations of
planetesimal disk which includes the self-gravity of the planetesimals 
with radial extent from 0.7 AU to 4.0 AU, including the
ice line.  The initial number of planetesimals is $\sim 82362$.
The simulations consist of two stages. The first stage simulations 
start with the planetesimals only to see the emergence of the 
runaway bodies.  Due to the increase of the solid surface density 
outside the ice line, growth of the planetesimals is accelerated.  Thus,
the runaway bodies form not only in the inner edge of the disk,
but also at just outside the ice line.  
In the second stage simulations, 
we replaced the runaway bodies outside the ice line 
with protoplanets of mass $m_{\rm proto}(\sim 0.1 M_{\oplus})$.
We assumed that the planetesimals within $5r_{\rm hill}$  
of $m_{\rm proto}$ are eaten and that a gap 
is formed in the planetesimal disk.
We restarted the simulation with two protoplanets and
the planetesimal disk from 0.7 AU to 4.0 AU.
The outer protoplanet moves outward and the inner protoplanet
moves inward due to PDM if necessary conditions are satisfied.
\subsection{Effect of Self gravity on Criteria of ML14}
ML14 investigated PDM using test particles.
They derived criteria for PDM to continue.
In this study, we carry out $N$-body simulations 
which include the self gravity of the planetesimals and 
investigate the criteria of PDM.
The inclusion of the self gravity of the planetesimals resulted in
the increase of random velocity of the planetesimals due to viscous
stirring.
This increase of the random velocity 
was not included in ML14.
We investigated whether the criteria of ML14 are satisfied or not in our
simulations.
The amount of planetesimals in front of the outgoing protoplanet
clearly affects the migration speed but in our simulations, 
criterion (1) was always satisfied in our simulations, 
and the criterion may be necessary.
For the mass resolution criterion, 
the criterion was satisfied in our simulations.
Hence, we conclude that the criterion (2) may be necessary as well.
The random velocity of the planetesimals increased in our simulations
due to the viscous stirring.  Since the migration stopped as the
eccentricity of the planetesimals exceeded $5-6h$, we conclude that
the criterion (3), which is the eccentricity criterion, determines whether 
PDM continues or not.

\subsection{Effect of Gas and Fragmentation}
In reality, 
the planetesimal accretion proceeds in the gas disk.
The presence of the gas nebula may be the key
to how PDM works.
We have shown that the stirring among the planetesimals
increase their random velocity which eventually
halts the outward migration.
Gas drag should keep the random velocity of the planetesimals small.
When a gas disk with several times MMSN is considered, 
the timescale for the gas drag to decrease $e$ and $i$ of 
planetesimals becomes shorter than the timescale of PDM.
Hence, PDM and its stopping may be naturally explained by
planetesimal-planetesimal gravitational interaction 
and the gas drag.  
The simulations which include the gas drag will be shown in the 
forthcoming paper.
When the gas disk is considered, type-I migration should be effective as well
(e.g. Tanaka et al. 2002).
Since the timescale for the type-I migration is comparable to
that of PDM when $0.1 M_\oplus$ is considered,  
$N$-body simulations including the effect of type-I migration
is also necessary to understand which one dominates for the outer protoplanet.
Depending on the mass density of the gas, PDM can actually be more 
effective than type-I migration for objects less than a few $M_\oplus$.
Details will be shown in the next paper.
PDM of the inner protoplanet stops at the ice line because of the decrease
of the solid surface density at the inner region of the ice line.
However, type-I migration
can carry the protoplanet to terrestrial planet region.

In addition, 
PDM process should be affected by the fragments formed
due to the collisions.
If fragmentation is considered, many small bodies should form.
The eccentricity of the fragments may be damped with short timescale
due to the gas drag. 
Hence, if the gas drag and fragmentation are considered at the same time, 
the number of bodies 
with small random velocity surrounding the migrating 
protoplanet should increase.
In the forthcoming paper, we will study the PDM within the gas disk.
We will include the fragmentation in our $N$-body simulation 
to consider more realistic model in forthcoming paper as well.

If the fragmentation and effect from the gas disk are included,
our artificial treatment that increased the mass of the runaway 
bodies between the first stage and the second stage of the simulation
can be replaced by more realistic continuous growth of the planetesimals
to the protoplanets.
We plan to perform such simulations soon.

\acknowledgements
The results were obtained by using
the K computer at the RIKEN Advanced Institute for Computational
Science (Proposal number hpci130026).  This work was supported in part
by MEXT SPIRE Field 5 “The origin of matter and the universe” and
JICFuS.

        
\clearpage
\begin{table}
\begin{center}
{\bf Table 1}
{\bf Model List}
\end{center}
  \caption{Summary of first stage simulations and
    second stage simulations.}\label{tab:first}
  \begin{center}
    \begin{tabular}{lll}
      1st stage simulation & 2nd stage simulation\\
      \hline
      run & run & $m_{\rm proto}(M_\oplus)$ \\
      \hline
      S1a & S2aM0.1 & 0.1 \\
          & S2aM0.3 & 0.3 \\
          & S2aM0.03 & 0.03 \\
      S1b & S2bM0.1 & 0.1 \\
      \hline
    \end{tabular}
  \end{center}
\end{table}

\clearpage
\begin{figure}
\begin{center}
 \includegraphics[scale=0.7]{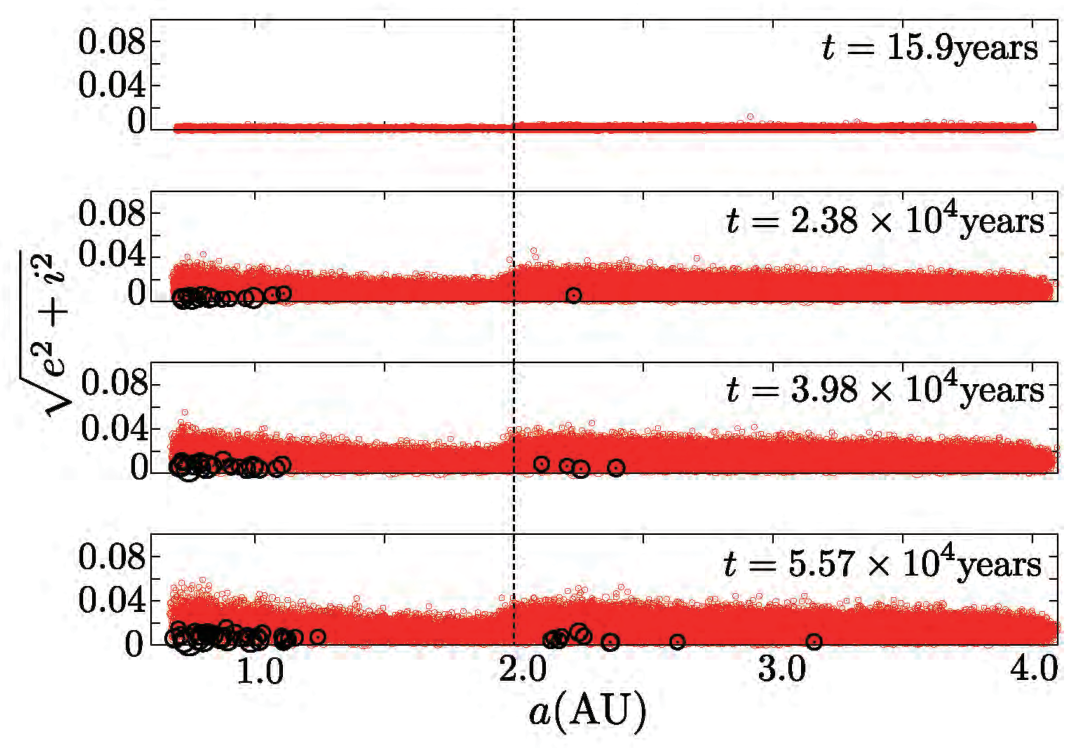}
\end{center}
\caption{The snapshots from run S1a on $a-\sqrt{e^2+i^2}$ plane.
The black circles are planetesimals with mass of more than ten 
times the initial mass.
Dashed line at $2$AU corresponds to the ice line.}
\label{fig.sim_1_e_a}
\end{figure}

\clearpage
\begin{figure}
\begin{center}
 \includegraphics[scale=0.7]{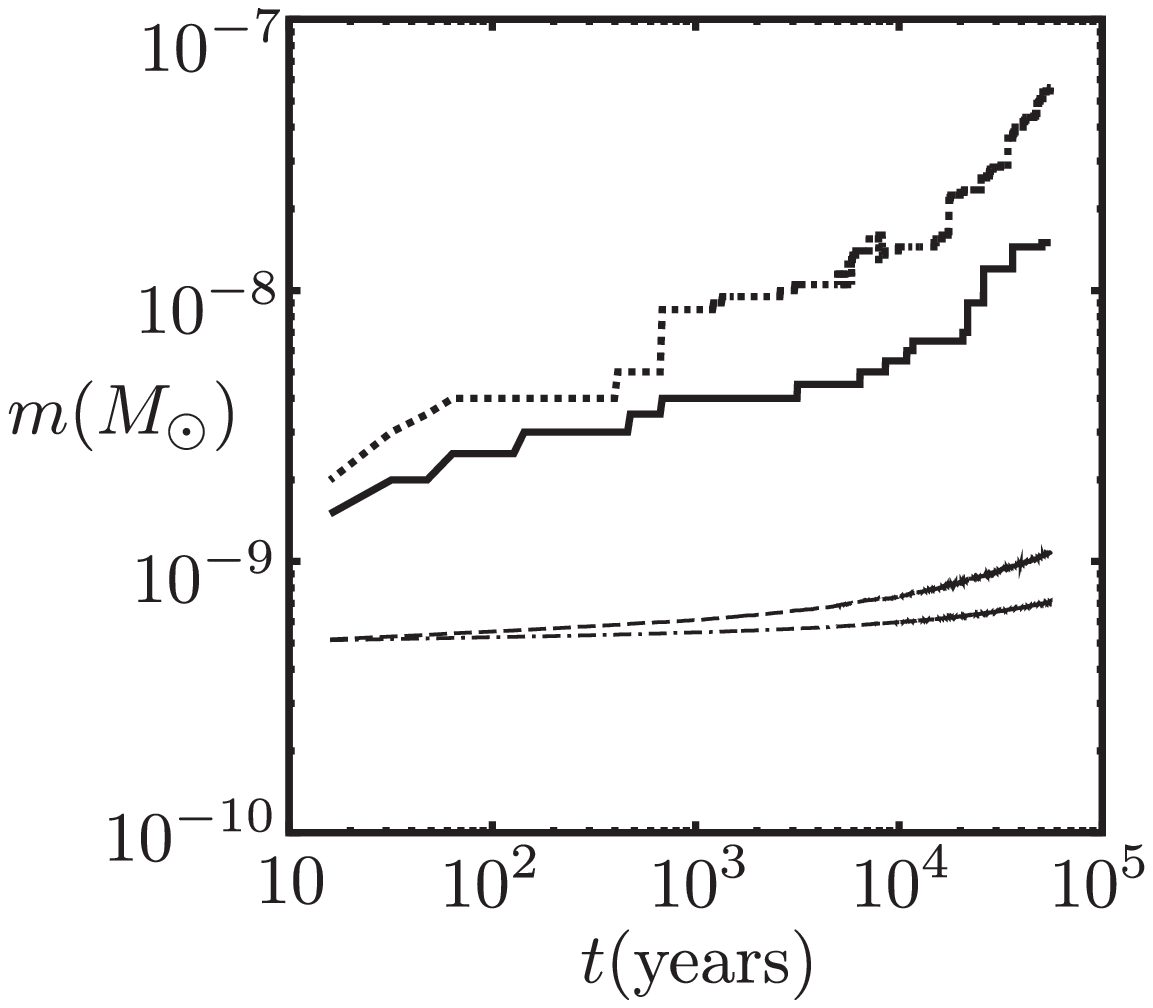}
\end{center}
\caption{Mass evolution of planetesimals in run S1a. 
The largest mass within the 
region from 0.7 to 1.0 is plotted in the dotted curve, and that of the 
region from 2.0 to 2.5 AU
is plotted in the solid curve.  The average mass of the planetesimals other
than the largest one is plotted in dashed curve (0.7-1.0AU) and 
dotted-dashed curve (2-2.5AU), respectively.}
\label{fig.RUNAWAY.INNER.OUTER}
\end{figure}

\clearpage
\begin{figure}
\begin{center}
 \includegraphics[scale=0.7]{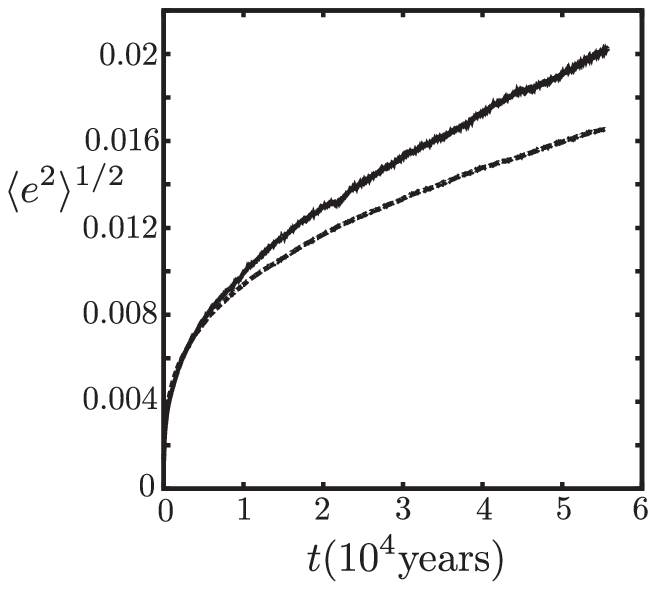}
\end{center}
\caption{Time evolution of the RMS eccentricity of the planetesimals 
within 0.7 - 1.0 AU (solid line) and that of those within 2.0 - 2.5 AU
(dotted line) in run S1a.}
\label{fig.ECC_INNER_OUTER}
\end{figure}

\clearpage
\begin{figure}
\begin{center}
 \includegraphics[scale=0.7]{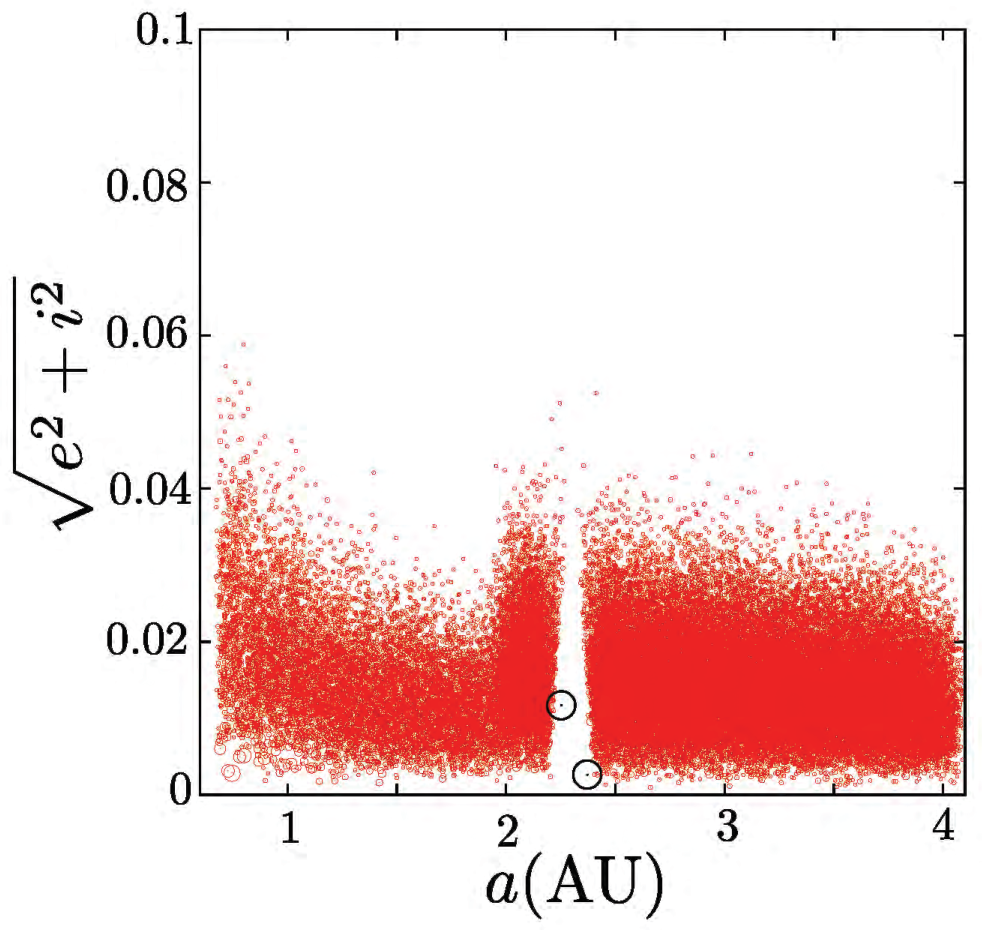}
\end{center}
\caption{The initial condition of run S2aM0.1.
Large black circles indicate the two protoplanets.}
\label{fig.sim_2_initial_e_a}
\end{figure}

\clearpage

\begin{figure}
\begin{center}
 \includegraphics[scale=0.7]{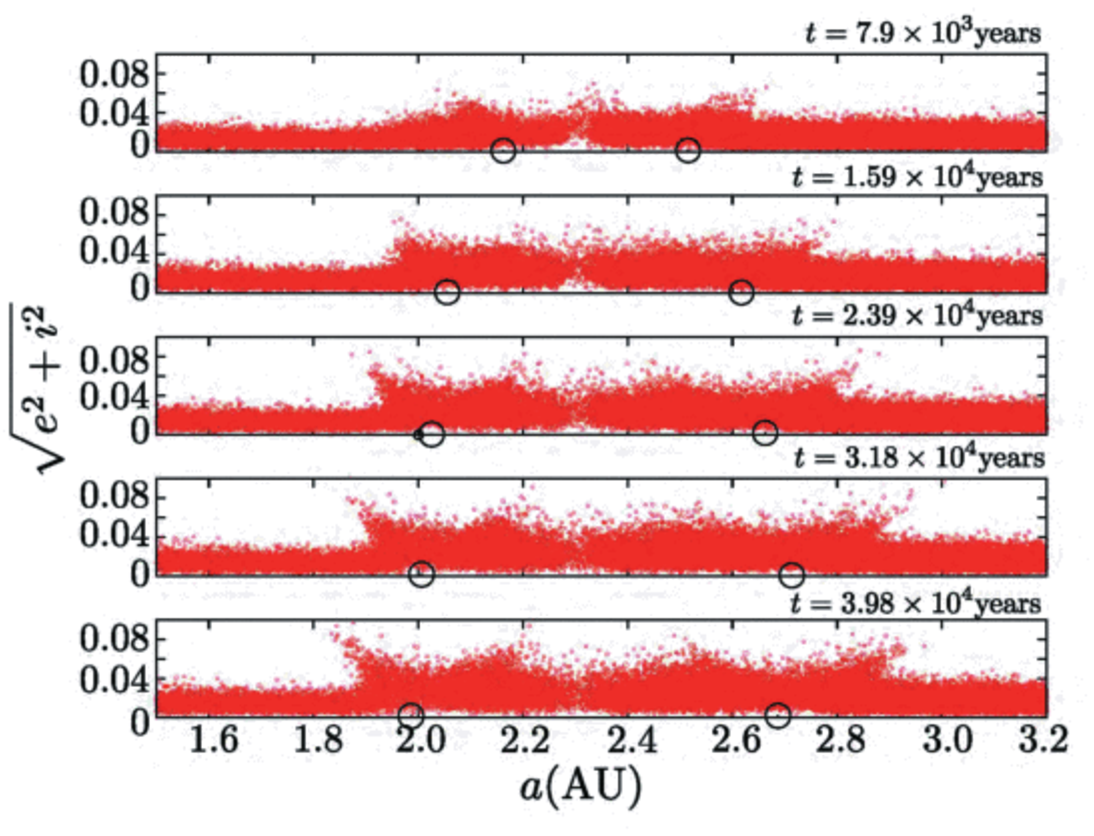}
\end{center}
\caption{Snapshots from run S2aM0.1 on the plane of $a$ and 
$\sqrt{e^2+i^2}$.
The protoplanets with initial mass of $m_{\rm proto}$
(in this case $0.1M_\oplus$) are shown in black circles. 
The size of the circles corresponds to the mass of the planetesimals
and the protoplanets.
}
\label{fig.sim_2_e_a}
\end{figure}

\clearpage
\begin{figure}
\begin{center}
 \includegraphics[scale=0.7]{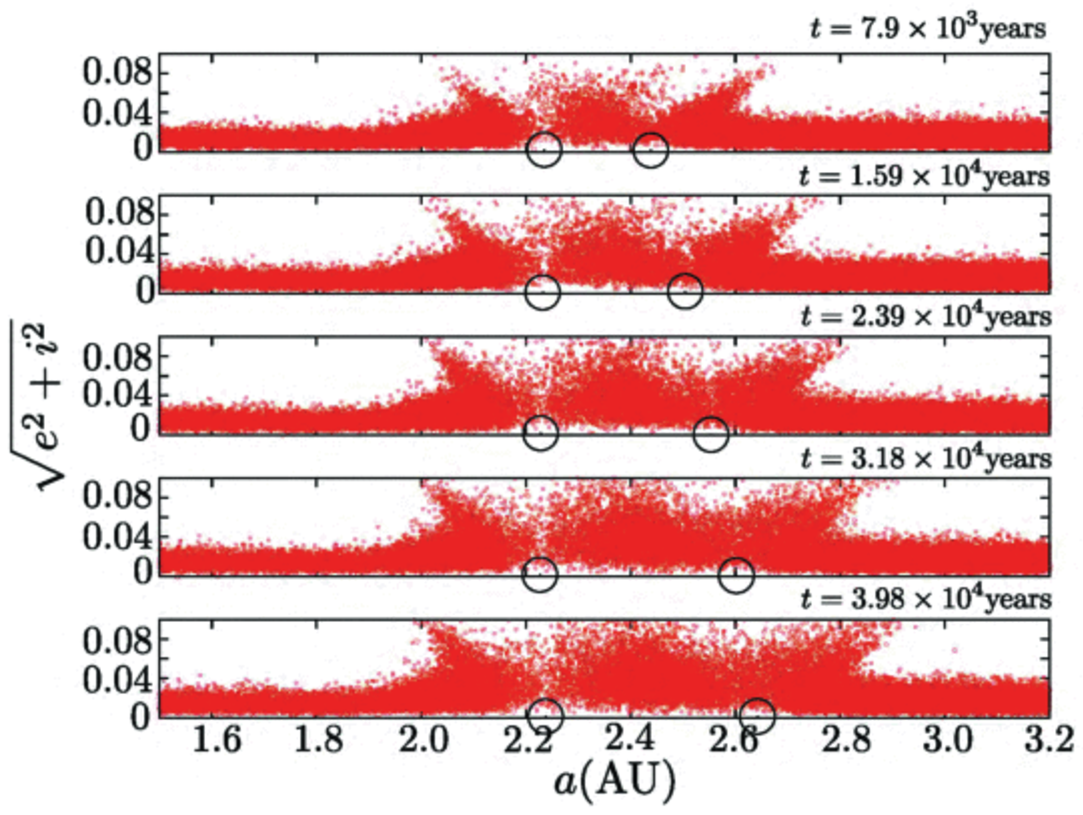}
\end{center}
\caption{Same as figure \ref{fig.sim_2_e_a} but for run S2aM0.3.}
\label{fig.sim_4_e_a}
\end{figure}

\clearpage

\begin{figure}
\begin{center}
 \includegraphics[scale=0.7]{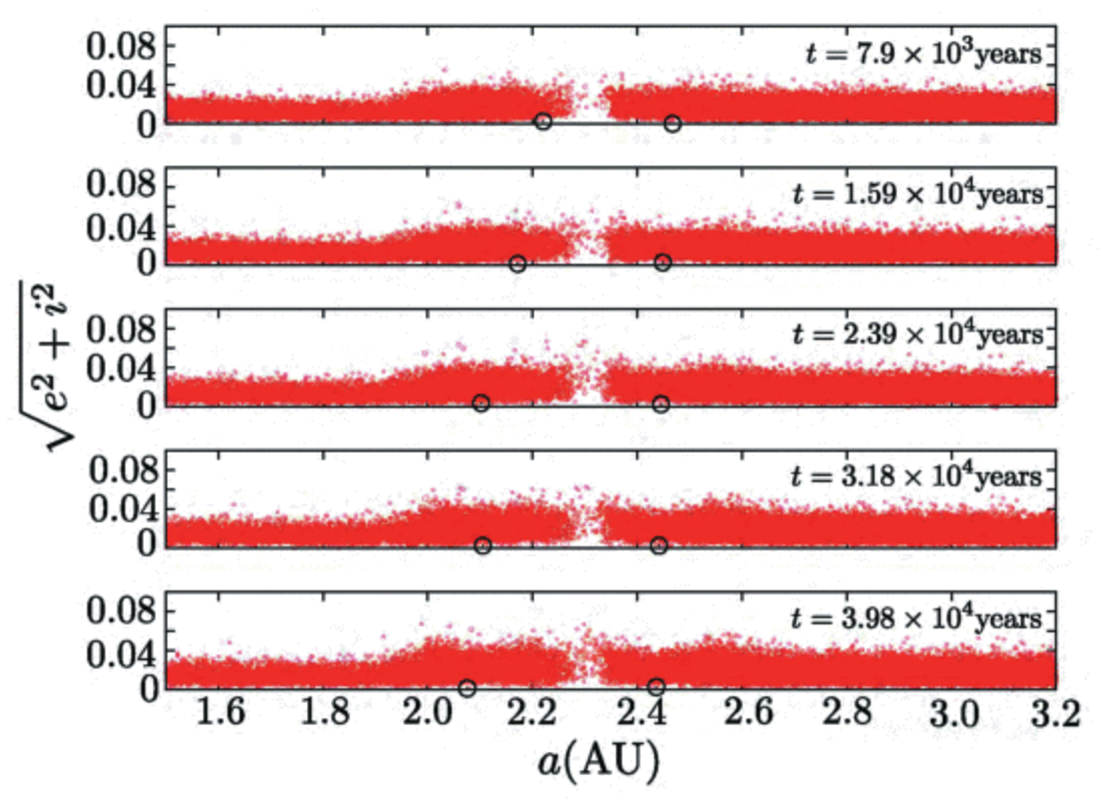}
\end{center}
\caption{Same as figure \ref{fig.sim_2_e_a} but for run S2aM0.03.}
\label{fig.sim_3_e_a}
\end{figure}

\begin{figure}
\begin{center}
 \includegraphics[scale=0.5]{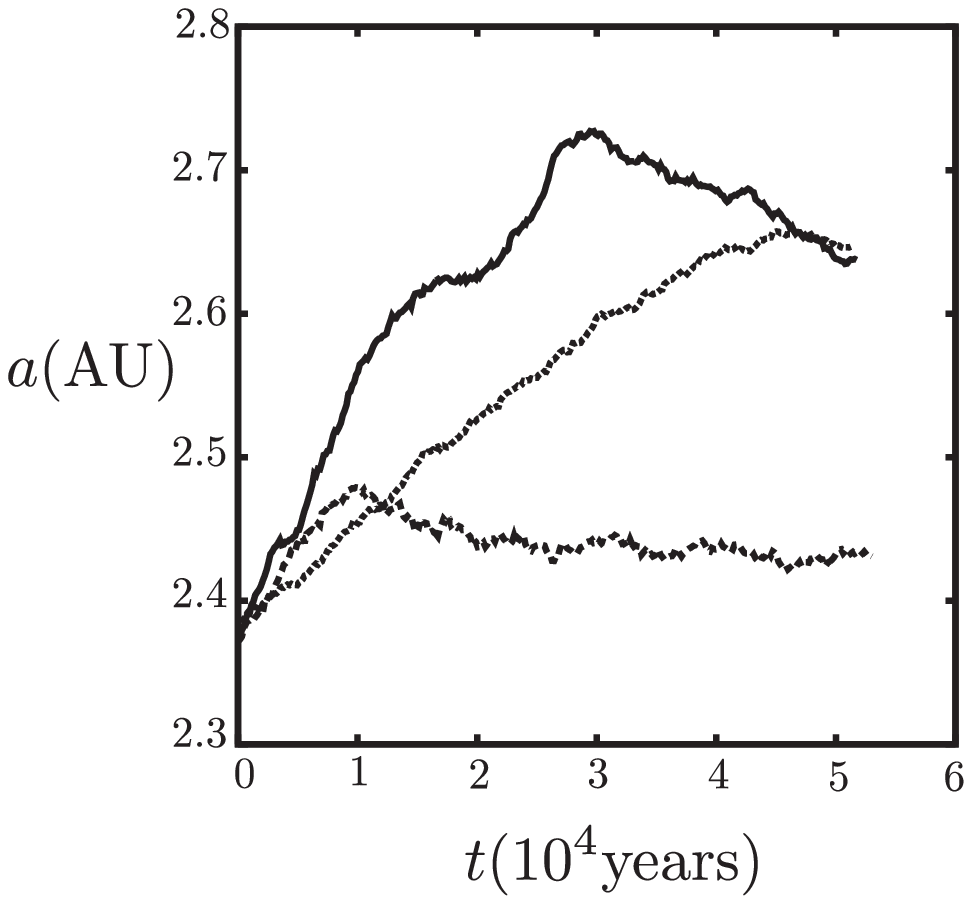}
\end{center}
\caption{Radial migration  of the outer protoplanets in runs S2aM0.03
(dashed curve), S2aM0.1(solid curve) and S2aM0.3(dotted curve).  
The horizontal axis is time in years and
the vertical axis is the semimajor axis of the protoplanet.}
\label{fig.sim_2_sa_out}
\end{figure}

\begin{figure}
\begin{center}
 \includegraphics[scale=0.5]{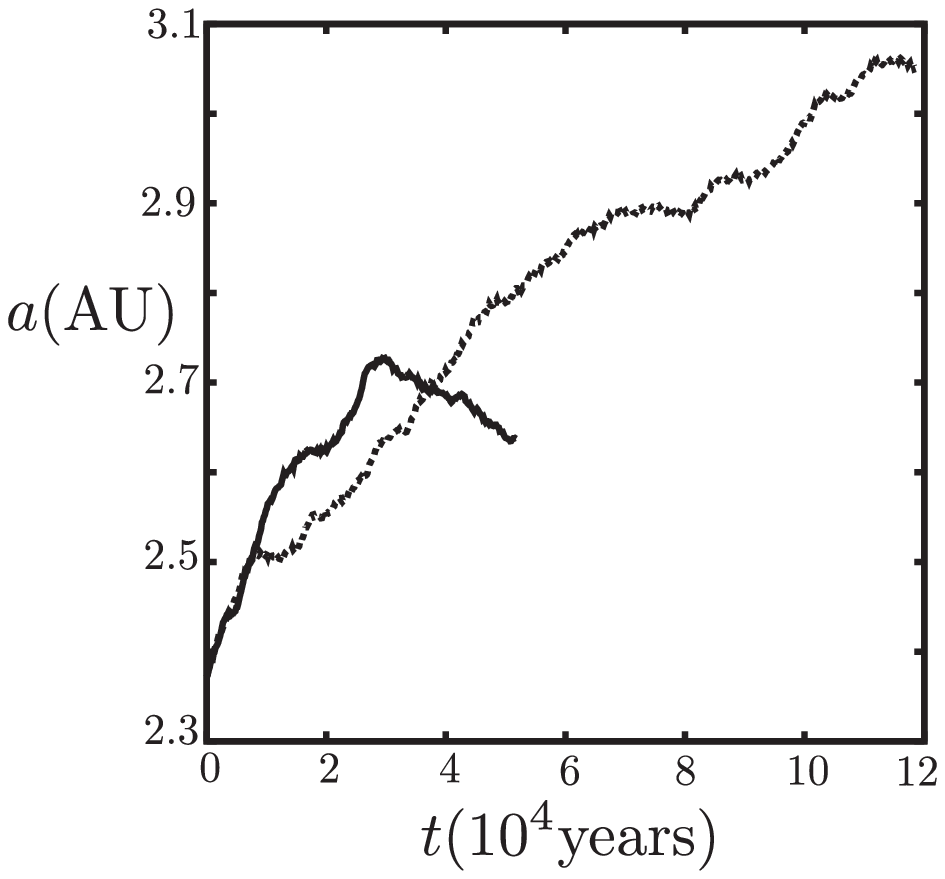}
\end{center}
\caption{Same as figure \ref{fig.sim_2_sa_out} but for run S2aM0.1
(solid curve) and S2bM0.1(dotted curve).}
\label{fig.sim2_evo_2sample}
\end{figure}


\clearpage
\begin{figure}
\begin{center}
 \includegraphics[scale=0.5]{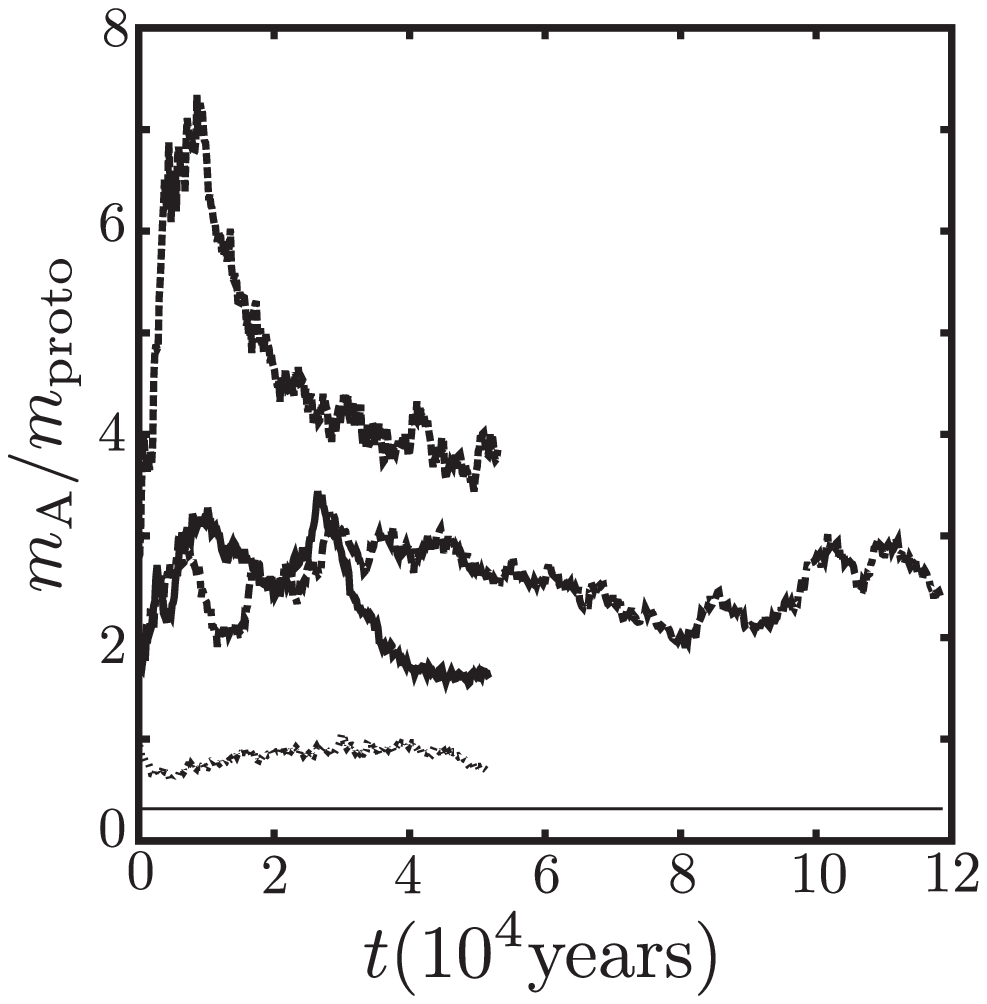}
\end{center}
\caption{Time evolution of the total planetesimal mass normalized by
the mass of the protoplanet within $5 r_{\rm hill}$
outside the protoplanet 
in runs S2aM0.1(solid curve), S2aM0.03(dashed curve), 
S2aM0.3(dotted curve) and S2bM0.1(long, dot-dashed curve).
The thin line is the line with {\bf $m_{\rm A}/m_{\rm proto}=1/3$} 
which shows the criterion (1) of ML14.
}
\label{fig.sim2_proto_mass.new}
\end{figure}

\clearpage

\begin{figure}
\begin{center}
 \includegraphics[scale=0.5]{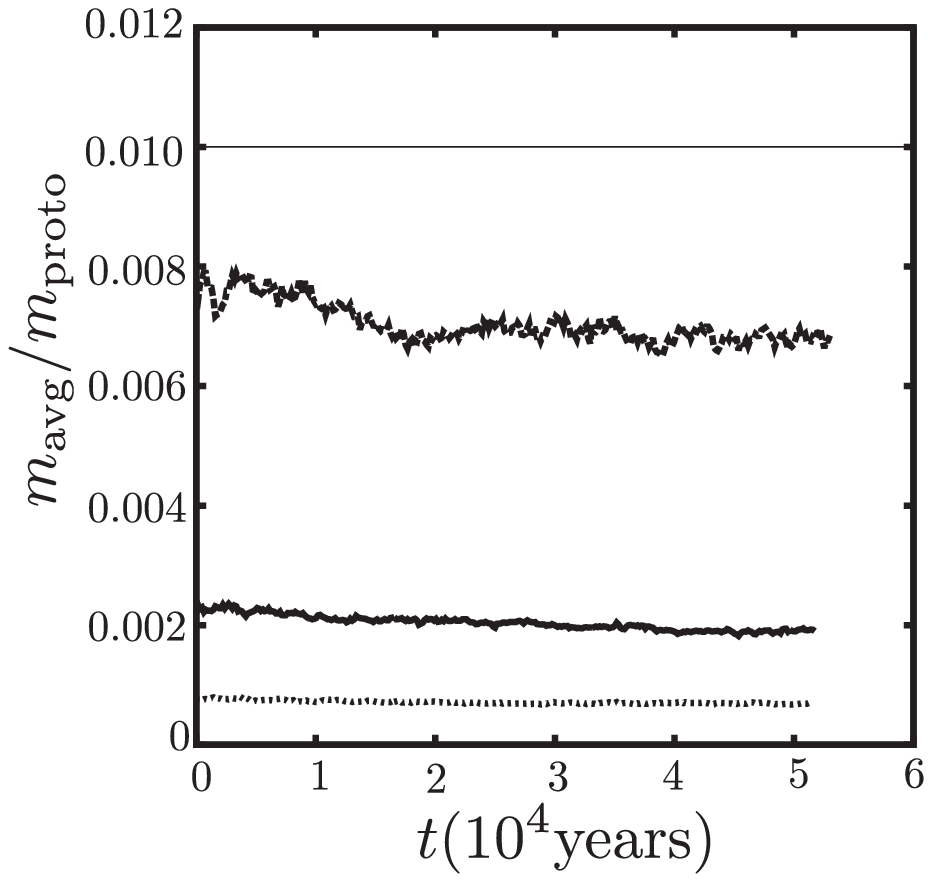}
\end{center}
\caption{Time evolution of the averaged mass of the planetesimals
within $5 r_{\rm hill}$ outside the protoplanet in S2aM0.1(solid curve),
S2aM0.03(dashed curve) and S2aM0.3(dotted curve).  The thin line shows the 
ratio where $m_{\rm avg}/m_{\rm proto}=0.01$, which is the line for 
criterion (2) in ML14.}
\label{fig.sim2_proto_avg.new}
\end{figure}

\clearpage

\begin{figure}
\begin{center}
 \includegraphics[scale=0.5]{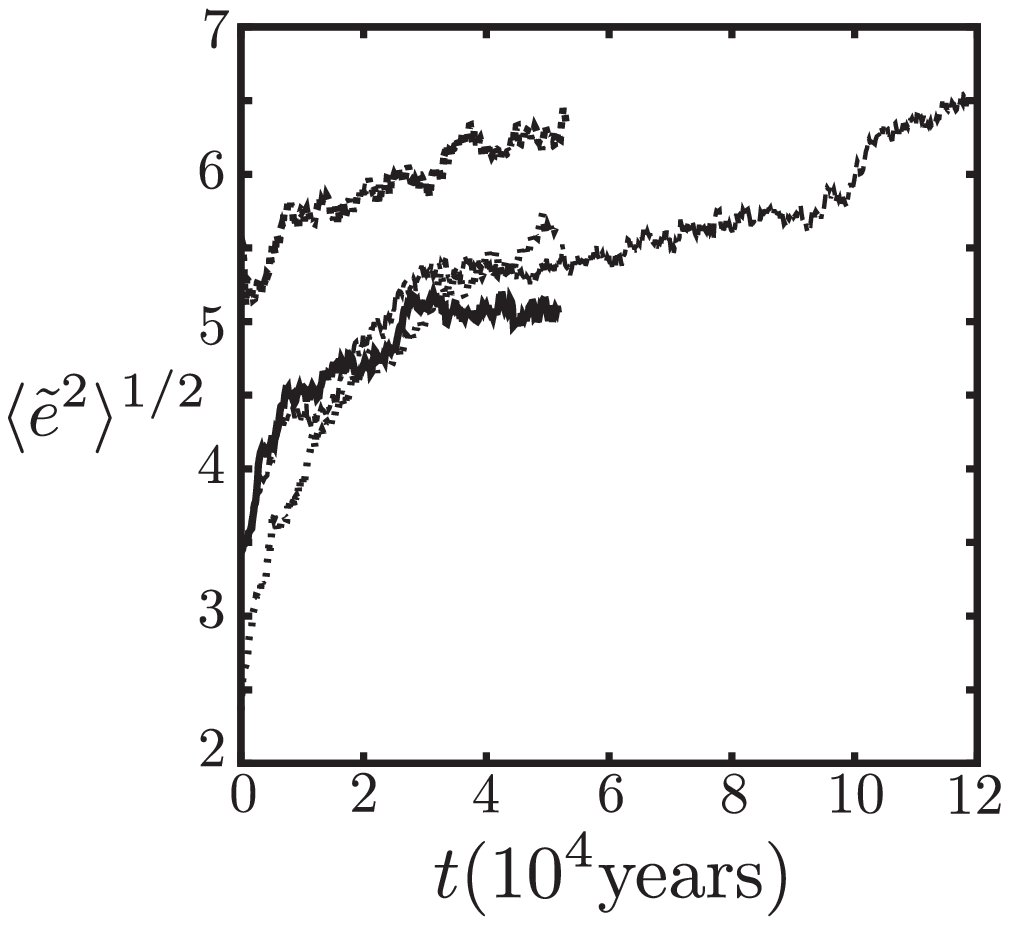}
\end{center}
\caption{Time evolution of the RMS eccentricity of the planetesimals
within $5 r_{\rm hill}$ outside the protoplanet in runs S2aM0.1(solid curve),
S2aM0.03(dashed curve), S2aM0.3(dotted curve) and S2bM0.1(dot-dashed curve).
}
\label{fig.eccevo_all.new}
\end{figure}



\clearpage


\begin{thebibliography}{}
\bibitem [Agnor et al. (1999)]{Agnoretal1999}
Agnor C.B., R.M. Canup \& H.F. Levison., 1999,
Icarus, 142, 219-237


\bibitem [Carter et al. (2015)]{Carteretal2015}
Carter, P. J., Z. M. Leinhardt, T. Elliott, M. J. Warter and S. T. Stewart, 2015,
ApJ(accepted).

%

\bibitem [Hayashi et al. (1985)]{Hayashietal1985}
Hayashi, C., K. Nakazawa. \& Y. Nakagawa., 1985,
Protostars and planets II, University of Arizona Press, 1100-1153.


\bibitem [Ida \& Makino (1992)]{IdaMakino1992}
Ida, S. \& Makino, J., 1992,
Icarus, 96,107-120.

\bibitem [Ida \& Makino (1993)]{IdaMakino1993}
Ida, S. \& Makino, J., 1993,
Icarus, 106,210-227.

\bibitem [Ida (1990)]{Ida1990}
Ida, S.,  1990,
Icarus, 88, 129-145.

\bibitem [Ida et al. (2000)]{Idaetal2000}
Ida, S., G. Bryden, D.N.C. Lin, \& H. Tanaka., 2000,
ApJ, 534, 428-445.

\bibitem [Kirsch et al. (2009)]{Kirschetal2009}
Kirsh, D. R., M. Duncan, R. Brasser \& H. L. Levison., 2009,
Icarus, 199, 197-209.
%
\bibitem [Kokubo \& Ida (1996)]{KokuboIda1996}
Kokubo, E., \& S. Ida., 1996,
Icarus, 123,180-191.

%
\bibitem [Kokubo \& Ida (1998)]{KokuboIda1998}
Kokubo, E., \& S. Ida., 1998,
Icarus, 131, 171-178.


\bibitem [Kokubo \& Ida (2002)]{KokuboIda2002}
Kokubo, E., \& S. Ida., 2002,
Icarus, 581, 666-680.

\bibitem [Kominami \& Ida (2002)]{KominamiIda2002}
Kominami, J \& S. Ida, 2002,
Icarus, 157, 43-56.
%
\bibitem [Kominami et al. (2016)]{Kominamietal2015}
Kominami, J.D., H. Daisaka, K. Nitadori \& J. Makino, 2016,
in prep.

%
\bibitem [Makino \& Aarseth (1992)]{MakinoAarseth1992}
Makino, J., \& S. J. Aarseth, 1992,
PASJ, 44,141-151.


\bibitem [Minton \& Levison (2014)]{MintonLevison2014}
Minton, D.A. \& H.F. Levison, 2014,
Icarus, 232, 118-132.

\bibitem [Morbidelli et al. (2012)]{Morbidellietal2012}
Morbidelli, A., J.I. Lunine, D.P. O'Brien, S.N. Raymond \& K.J. Walsh, 2012,
Annal Review of Earth and Planetary Sciences, 40, 251-275.

\bibitem [Nakazawa \& Ida (1988)]{NakazawaIda1988}
Nakazawa K. \& S. Ida, 1988,
Prog.Theoy.Phys.Suppl., 96, 167-174.


\bibitem [Nitadori et al. (2006)]{Nitadorietal2006}
Nitadori, K., J. Makino \& G. Abe, 2006,
arXiv:astro-ph/0606105v2.

%

\bibitem [Oka et al. (2011)]{Okaetal2011}
Oka, A., T. Nakamoto \& S. Ida, 2011,
ApJ, 738, Issue 2, article id. 141, 11 pp.

\bibitem [Ormel et al. (2012)]{Ormeletal2012}
Ormel, C.W., S. Ida \& H. Tanaka, 2012,
Astrophys J., 758, 80, (17pp)

\bibitem [Raymond et al. (2012)]{Raymondetal2012}
Raymond, S. N., P. J. Armitage, A. Moro-Martín, M. Booth, M. C. Wyatt,  J. C. Armstrong,  A. M. Mandell, F. Selsis \& A. A. West, 2012,
A\&A, 541 et.A11,25pp

%
\bibitem [Tanaka et al. (2002)]{Tanakaetal2002}
Tanaka, H., T. Takeuchi \& W. Ward, 2002,
ApJ, 565,1257-1274.

                                                                              

\bibitem [Walsh et al. (2011)]{Walshetal2011}
Walsh, K. J., A. Morbidelli, S. N. Raymond., D. P. O'brien \& A. M. Mandell, 
2011,
Nature, Volume 475, Issue 7355, pp. 206-209.

\bibitem [Wetherill \& Stewart (1989)]{WetherillStewart1989}
Wetherill G. W. \& G. R. Stewart, 1989,
Icarus, 77, 330-357.

\end{thebibliography}
\end{document}